\documentclass[twocolumn,showpacs,preprintnumbers,amsmath,amssymb]{revtex4}

\usepackage{color}
\usepackage{graphicx}
\usepackage{comment}

\begin{document}

\title{Zero-Temperature Configurations of Short Odd-Numbered Classical Spin Chains with Bilinear and Biquadratic Exchange Interactions}

\author{N. P. Konstantinidis}
\affiliation{Max Planck Institut f\"ur Physik komplexer Systeme, 01187 Dresden, Germany}

\date{\today}

\begin{abstract}
The lowest energy configurations of short odd open chains with classical spins are determined for antiferromagnetic bilinear and biquadratic nearest-neighbor exchange interactions. The zero field residual magnetization generates differences with the magnetic behavior of even chains, as the odd chain is like a small magnet for weak magnetic fields. The lowest energy configuration is calculated as a function of the total magnetization $M$, even for $M$ less than the zero field residual magnetization. Analytic expressions and their proofs are provided for the threshold magnetic field needed to drive the system away from the antiferromagnetic configuration and the spin polar angles in its vicinity, when the biquadratic interaction is relatively weak. They are also given for the saturation magnetic field and the spin polar angles close to it. Finally, an analytic expression along with its proof is given for the maximum magnetization in zero magnetic field for stronger biquadratic interaction, where the lowest energy configuration is highly degenerate.
\end{abstract}


\maketitle

\section{Introduction}
\label{sec:Introduction}
In recent years important advances in synthetic chemistry have led to the production of molecules that have interesting magnetic properties and could potentially provide the building blocks for quantum computers and memory devices. This class of molecules has been coined molecular nanomagnets \cite{Gatteschi06}. Another route for the production of small entities with interesting magnetic properties that has recently seen intense activity is the artificial engineering of molecular nanomagnets, magnetic clusters and arrays of magnetic adatoms, which are fabricated directly on surfaces with Scanning Tunneling Microscopy (STM) \cite{Eigler90,Hirjibehedin06,Meier08}. STM is also used to measure their magnetic properties. These entities have brought forward the need to understand thoroughly magnetic interactions between a few magnetic centers and their collective magnetic behavior \cite{Furrer13,NPK11}, especially since their properties are expected to be different in comparison with systems that contain many magnetic centers and are relatively close to the thermodynamic limit \cite{Auerbach98,Fazekas99}.

Perhaps the simplest non-trivial magnetic entity is an open linear chain with equivalent magnetic centers and bilinear antiferromagnetic (AFM) exchange interactions between them. Its quantum-mechanical behavior as a function of the spin magnitude of the magnetic centers $s$ has been the subject of extensive investigation. For $s=1/2$ and nearest-neighbor interactions the model is integrable via the Bethe ansatz \cite{Bethe31}. The Haldane conjecture differentiates the magnetic properties of the infinite chain according to the parity of $s$ \cite{Haldane83-1,Haldane83-2}. For finite chains the boundary conditions play an important role \cite{Machens13,Baerwinkel00,Schnack10}. An open odd-membered chain is not frustrated, in contrast to its closed counterpart where frustration forces the spins in a non-collinear configuration at the classical level. In contrast the open chain, even though it lacks frustration, is more susceptible towards the edges to an external magnetic field due to the lack of translational symmetry. This has been demonstrated experimentally \cite{Micotti06,Furukawa08}, and as a result the odd open chain possesses interesting magnetic properties even at the classical level.
Even and odd short open chains show pronounced differences in their magnetic behavior. The lowest energy as a function of total spin $S$ depends on the parity of the chain \cite{Machens13}. At the classical level the lowest energy configuration of an even chain is always non-collinear in a field \cite{Rossler04,Rossler04-1}, while an odd chain changes from a ferrimagnetic to a non-collinear configuration at a threshold value of the field \cite{Lounis08,Politi09}. This is because in zero field nearest-neighbor spins are antiparallel, however the net magnetization of an odd chain is not zero due to the single uncompensated spin. Thus in small fields the chain is behaving like a small magnet that gains magnetic energy without having to break the AFM configuration. It takes a finite magnetic field to destroy the zero field AFM configuration, whereas an even chain immediately responds to a field.
The residual magnetization of classical AFM odd chains implies that only magnetizations higher than it are accessible with application of an external field. Still, there are configurations where the total magnetization is less than the residual. For each such magnetization there is a configuration that has minimum energy, which has been calculated numerically in Ref. \cite{Machens13}.

Consideration of magnetic models for classical spins provides a first estimate of the influence of connectivity and boundary conditions on magnetic properties, before quantum fluctuations are taken into account. For more complicated connectivities than the chain, or more generally for structures that are not bipartite or for bipartite structures with competing interactions, frustration plays an important role. Magnetization discontinuities have been found in frustrated clusters, where specific ranges of the total magnetization $M$ never include the lowest energy configuration in a magnetic field. Such is the case for fullerene clusters with icosahedral $I_h$ spatial symmetry, where the dependence of the spin polar angles on the field can also be non-monotonic like an even or odd chain \cite{NPK05,NPK07}, and the icosahedron which also has $I_h$ spatial symmetry \cite{Schroeder05,NPK15}. The classical approach is also important as it describes the quantum problem for relatively high quantum numbers \cite{Hirjibehedin06}. In the case of open chains this description is valid down to low $s \geq \frac{3}{2}$ \cite{Machens13}. This agreement has also been demonstrated for low $s$ for open and closed chains with an extra spin attached \cite{NPK13}.

A standard approach to analyze magnetic properties is to consider first models with bilinear exchange interactions between the system's constituents. The strongest interactions are between nearest-neighbors, but interactions more distant than nearest-neighbor can also be taken into account. Nevertheless it has been shown that higher-order exchange terms are often important for the analysis of experimental data \cite{Furrer11}. A biquadratic exchange interaction term is needed to explain the slight deviations from the Land\'e rule for a $Mn^{2+}$ dimer in a $CsMgBr_3$ single crystal \cite{Falk84}. It has also been found important for the explication of magnetic susceptibility data of quasiclassical one-dimensional magnetic systems \cite{Gaulin86}. The biquadratic exchange interaction was derived microscopically by Anderson \cite{Anderson63}. For spins $s$ the most general interaction that is rotationally invariant is a polynomial of degree $2s$ in $\vec{s}_i \cdot \vec{s}_j$, where $\vec{s}_i$ and $\vec{s}_j$ are two of the spins of the system. Thus the biquadratic exchange is the next-order allowed term after the bilinear exchange interaction. Its inclusion in the Hamiltonian further complicates the problem. In contrast to the AFM bilinear exchange which favors antiparallel nearest-neighbor spins, the biquadratic interaction with a positive prefactor is minimized for perpendicular nearest-neighbors. The competition of the two terms determines the lowest energy configuration where nearest-neighbor spins are not necessarily antiparallel in zero field when the biquadratic interaction is strong enough. This means that the ground state configuration is not necessarily coplanar, introducing a degeneracy for the ground state whose degenerate manifold corresponds to a continuous range of values of the residual magnetization. The lowest energy configurations for magnetizations lower than the minimum magnetization of the ground state manifold have not been determined up to now in the literature.

The model with an AFM bilinear term and biquadratic exchange has been solved with the transfer matrix method in the thermodynamic limit in one dimension for arbitrary temperature \cite{Thorpe72}, where it was shown that when the strength of the biquadratic interactions becomes equal to half the strength of the bilinear interactions the lowest energy configuration changes from an AFM one to a spiral. The same applies for ferromagnetic bilinear exchange, with the lowest energy configuration for weak biquadratic interaction being ferromagnetic. An extension was made to include AFM next nearest-neighbor interactions, which eventually generate an up-up-down-down ground state when their strength becomes one-half of the ferromagnetic nearest-neighbor interaction and the biquadratic interaction has a negative prefactor \cite{Kaplan09}. The model has also been considered in higher dimensions \cite{Hayden10,Kawamura07,Wenzel13}.

The magnetic properties of molecular nanomagnets depend subtly on the nature of the interaction between their magnetic constituents, especially in conjuction with the cluster's size, shape and symmetry. The magnetic properties of the AFM Heisenberg model on open chains were found to strongly depend on chain parity \cite{Machens13}. In this paper we add to the AFM bilinear the biquadratic exchange interaction with a positive prefactor, and consider the case of odd open chains. Inclusion of a magnetic field in the Hamiltonian allows for the determination of the lowest energy configuration for $M$ greater than the residual magnetization in the absence of a field. Furthermore, when the Zeeman term is replaced by a term proportional to $S^2$ the lowest energy configuration can be determined even in the case where $M$ is less than the zero field residual magnetization. The latter is 1 when the biquadratic interaction is less than half of the bilinear interaction and corresponds to a continuous range of values in the opposite case, whose maximum is analytically derived while the minimum is calculated numerically. The analytic expression and its derivation are provided for the threshold magnetic field above which the residual spin starts to respond when the biquadratic interaction is less than half the bilinear interaction, and the corresponding values of the spin polar angles are also analytically calculated in this limit. The same is done for the saturation magnetic field, with the corresponding formulas in this case being valid also for even chains.

The plan of the paper is as follows: in Sec. \ref{sec:Model} the model is presented along with the methods of calculation of the lowest energy configuration as a function of $M$. Section \ref{sec:generalanalyticresults} includes analytic expressions for the ground state configuration and its magnetization in the absence of a field, as well as the threshold and saturation field and the spin polar angles in their immediate vicinity. Section \ref{sec:exchangeonly} presents results for the lowest energy configuration as a function of $M$ for zero biquadratic interaction. Section \ref{sec:exchangewithbiquadratic} presents the corresponding results for non-zero biquadratic interaction, while Sec. \ref{sec:conclusions} provides the conclusions. Three appendices provide derivations of analytic expressions presented in the main text.

\section{Model}
\label{sec:Model}
The Hamiltonian of the bilinear-biquadratic exchange model for an open chain is
\begin{eqnarray}
H & = & \sum_{i=1}^{N-1} [ J \vec{s}_i \cdot \vec{s}_{i+1}  + J' ( \vec{s}_i \cdot \vec{s}_{i+1} )^2 ] - h \sum_{i=1}^N s_i^x .
\label{eq:Hchain}
\end{eqnarray}
$N$ is the number of spins, taken to be odd. $J$ is the strength of the bilinear and $J'$ of the biquadratic exchange interaction. Both of them are taken positive, with $J$ defining the unit of energy.
The magnetic field $\vec{h}$ is taken along the $\hat{x}$ direction. The spins $\vec{s}_i$ are classical unit vectors and each is defined by a polar and an azimuthal angle. Hamiltonian (\ref{eq:Hchain}) is minimized with respect to these angles \cite{Coffey92,NPK05,NPK07,NPK15}. All the lowest energy configurations in a non-zero field are found to be coplanar, so it suffices to express the corresponding spin directions in polar coordinates where the angles $\theta_i$ with $i=1,\dots,N$ fully determine the spin configuration. The bilinear exchange interaction favors antiparallel nearest-neighbor spins, while the biquadratic perpendicular. The biquadratic interaction is the square of the bilinear, therefore it is weaker in strength. Simultaneously the spins gain magnetic energy from the field, therefore the physics is determined by the competition of the three terms. As the field varies it singles out the lowest energy configurations for $M$ greater than the (maximum) value in the global (irrespective of $M$) ground state(s), and essentially one can think of the ground state for a given $M$ as being effected by the field. To calculate the lowest energy as a function of $M$ it is only needed to subtract the magnetic energy, thus results are presented as a function of $M$ and not $h$ and the lowest energies given as functions of $M$ do not take into account the magnetic energy. In the ground state spins symmetrically placed with respect to the center have the same polar angle.

To calculate the lowest energy configuration for $M$ less than the (minimum) value in the global ground state(s) the constraint for constant $S$ is added to Hamiltonian (\ref{eq:Hchain}). This is done by introducing a Lagrange multiplier $G$ that multiplies $S^2$, instead of the magnetic field energy. The new Hamiltonian is then minimized for $G>0$. When $G<0$ minimization of the Hamiltonian produces the ground state for magnetizations higher than the (maximum) residual. Since $G S^2=G (\sum_{i=1}^N s_i)^2=G(2\sum_{i>j}\vec{s}_i \cdot \vec{s}_j + N)$ it is
\begin{eqnarray}
H & = & \sum_{i=1}^{N-1} [ J \vec{s}_i \cdot \vec{s}_{i+1}  + J' ( \vec{s}_i \cdot \vec{s}_{i+1} )^2 ] + \nonumber \\ & & G(2\sum_{i>j}\vec{s}_i \cdot \vec{s}_j + N) .
\label{eq:Hchain2}
\end{eqnarray}
The $M=0$ lowest state is generated when $G \to \infty$. From a numerical point of view when $G$ is very large and consequently $M$ approaches zero loss of numerical precision precludes the very accurate calculation of the corresponding lowest energy configuration. In this case the accurate results for very small $M$ are extrapolated down to $M=0$. Comparing with the $N=3$ analytic solution, and judging by the consistency of the results as more points are included in the extrapolation for arbitrary $N$, the extrapolated results are very accurate. In addition, results for different $N$ are absolutely consistent with each other.
Similarly to the case of Hamiltonian (\ref{eq:Hchain}), all the lowest energy configurations for $G>0$ are found to be coplanar, and the energy as a function of $M$ is calculated by subtracting the term $GM^2$. In the lowest energy configurations for $M$ less than the (minimum) value in the global ground state(s) spins symmetrically placed with respect to the center have angles adding up to $2\pi$. For the smallest chain $N=3$ analytic results can be generated for the whole range of $M$ with minimization, and these will be presented in Secs \ref{sec:exchangeonly} and \ref{sec:exchangewithbiquadratic}. It is noted that all analytic results presented have been verified numerically for a wide variety of cases.

\section{General Analytic Results}
\label{sec:generalanalyticresults}

\subsection{Maximum Residual Magnetization of the Global Ground State Manifold for $J' > \frac{J}{2}$}
\label{subsec:minimummaximummagnetization}

In the absence of magnetic field Hamiltonian (\ref{eq:Hchain}) can be written as
\begin{eqnarray}
H_{h=0}=J'\sum_{i=1}^{N-1} (\vec{s}_i \cdot \vec{s}_{i+1} + \frac{J}{2J'})^2-(N-1)\frac{J^2}{4J'} .
\label{eq:Hchainzerofield}
\end{eqnarray}
Minimization of Hamiltonian (\ref{eq:Hchainzerofield}) gives the global lowest energy configuration, and requires minimization of the square. For $J' \leq \frac{J}{2}$ Hamiltonian (\ref{eq:Hchainzerofield}) is minimized by the AFM configuration, with $E_g=-(N-1)(J-J')$. For $J' > \frac{J}{2}$ the nearest-neighbor angle that minimizes Hamiltonian (\ref{eq:Hchainzerofield}) is $\theta_g=\arccos(-\frac{J}{2J'})$, with $E_g=-(N-1)\frac{J^2}{4J'}$ and $\theta_g$ decreases with increasing $J'$ from $\pi$ to $\frac{\pi}{2}$. While the spin configuration is well-defined in the AFM state with a residual magnetization $M_{g,J' \leq \frac{J}{2}}=1$, when $J' > \frac{J}{2}$ the only requirement is that the angle between any two neighboring spins is $\theta_g$. This is satisfied by a degenerate manifold of configurations that are in general non-coplanar \cite{Thorpe72,Wenzel13}.

\begin{figure}
\includegraphics[width=3.5in,height=2.8in]{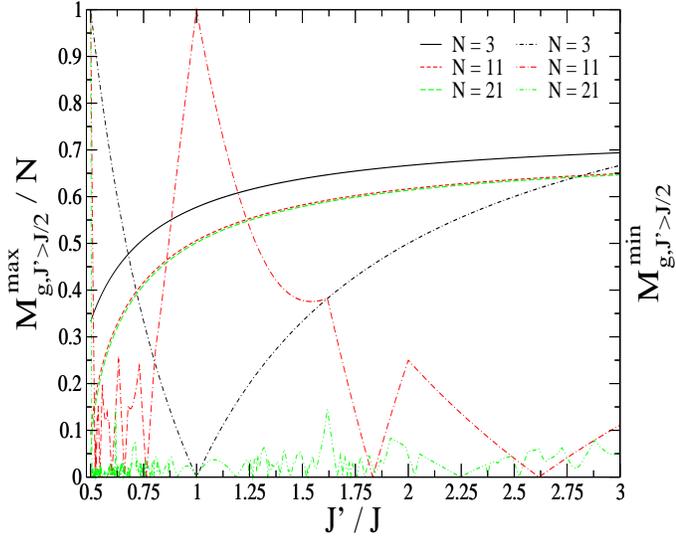}
\vspace{0pt}
\caption{Maximum reduced $M_{g,J'>J/2}^{max}/N$ (solid line: $N=3$, dashed line: $N=11$, long-dashed line: $N=21$), and minimum $M_{g,J'>J/2}^{min}$ (dot-dashed line: $N=3$, dot-long dashed line: $N=11$, double dot-dashed line: $N=21$) global ground state magnetization for open chains of different length $N$ as a function of $\frac{J'}{J}>\frac{1}{2}$. It is $M_{g,J' \leq J/2}=1$. The maximum reduced magnetization is given by Eq. (\ref{eq:resspinmax}) and does not change significantly for larger $N$ and stronger $J'$.
}
\label{fig:redmagngrstateminmax}
\end{figure}

The configuration of the ground state manifold of Hamiltonian (\ref{eq:Hchainzerofield}) with maximum total magnetization $M_{g,J'>\frac{J}{2}}^{max}$ is selected by an infinitesimal magnetic field. To maximize the total magnetization (and the magnetic energy) all spins are coplanar. In particular, $\frac{N+1}{2}$ spins are at an angle
\begin{eqnarray}
\theta_0 = \arctan\frac{\sqrt{1-\frac{J^2}{4J'^2}}}{\frac{N+1}{N-1} - \frac{J}{2J'}}
\end{eqnarray}
with the field, while the rest form an angle $\theta_g$ with their neighbors and $\theta_g-\theta_0$ with the field (see App. \ref{appendix:limitingmagnetizationfromabove}). It is noted that $\theta_0$ does not change significantly for larger $N$ and stronger $J'$. The maximum total magnetization is
\begin{eqnarray}
M_{g,J' > \frac{J}{2}}^{max} & = & \frac{1}{2\sqrt{1+\frac{1-\frac{J^2}{4J'^2}}{(\frac{N+1}{N-1}-\frac{J}{2J'})^2}}} [ N+1 + \nonumber \\ & & (N-1) (-\frac{J}{2J'} + \frac{1-\frac{J^2}{4J'^2}}{\frac{N+1}{N-1} - \frac{J}{2J'}}) ]
\label{eq:resspinmax}
\end{eqnarray}
and is plotted in Fig. \ref{fig:redmagngrstateminmax}. The limiting forms are $\frac{M_{g,J' > \frac{J}{2},N \to \infty}^{max}}{N} = \sqrt{\frac{1-\frac{J}{2J'}}{2}}$ and $\frac{M_{g,J' \to \infty}^{max}}{N} \to \sqrt{\frac{N^2+1}{2N^2}}$.


When $J' > \frac{J}{2}$ the minimum magnetization configuration of the ground state manifold is also coplanar and has the spins spread out as much as possible. Its total magnetization $M_{g,J' > \frac{J}{2}}^{min}$ is determined numerically and is plotted in Fig. \ref{fig:redmagngrstateminmax}.

\subsection{Threshold Magnetic Field for $J' \leq \frac{J}{2}$}
\label{subsec:thresholdfield}
When $J' \leq \frac{J}{2}$ the lowest energy configuration in the abscence of a field is AFM, and remains so for small non-zero fields. Unlike an even chain which has no zero field residual magnetization, the odd chain gains immediately magnetic energy in a field while simultaneously allowing the exchange energy not to change. Thus it takes a finite magnetic field to destroy the AFM configuration in order to increase the magnetic energy even more at the expense of the exchange energy. This threshold value of the field (given in \cite{Politi09} for $J'=0$) is
\begin{eqnarray}
h_t=2(J-2J')\sin\frac{\pi}{2N}
\label{eq:thresholdfield}
\end{eqnarray}
(see App. \ref{appendix:thresholdfield}). $h_t$ is proportional to $J$ but decreases with $J'$, and vanishes for an infinite chain where the parity of the chain makes no difference. The spin polar angles right above $h_t$ are
\begin{eqnarray}
\theta_i & = & \big\{ \sin\{\frac{\pi}{2N}(i-1)\} + \cos(\frac{\pi}{2N}i) \big\} \theta_0 \textrm {, i odd} \nonumber \\
\theta_i & = & \pi + \big\{ \sin(\frac{\pi}{2N}i) + \cos\{\frac{\pi}{2N}(i-1)\} \big\} \theta_0 \textrm {, i even}
\end{eqnarray}
with $i=1,\dots,N$ and $\theta_0$ a very small parameter that goes to 0 at the AFM configuration (see App. \ref{appendix:thresholdfield}). This is in contrast to an even chain that has been diverted from the AFM configuration by an infinitesimal field, where the deviation from the AFM configuration depends linearly on the site index \cite{Machens13}. The deviation from the AFM configuration is stronger for even spins and increases going towards the center of the chain. For large $N$ the magnitude of the deviation is proportional to $1 + \frac{\pi}{2N} (i-1)$ for odd and $1 + \frac{\pi}{2N} i$ for even spins not very far from the edges, thus the difference in the deviation from the AFM configuration between successive spins is proportional to $\frac{\pi}{2N}$ when going towards the center. It is constant like the one of the even chain, showing again than for an infinite chain its parity plays no role.

\subsection{Saturation Magnetic Field}
\label{subsec:saturationmagneticfield}
The saturation magnetic field required to reach the ferromagnetic configuration is
\begin{eqnarray}
h_s=2(J+2J')(1+\cos\frac{\pi}{N})
\label{eq:saturationfield}
\end{eqnarray}
(see App. \ref{appendix:saturationfield}). This formula is also valid for even chains, as no assumption for the parity is made in its derivation. For an infinite chain lim$_{N \to \infty}h_s=4(J+2J')$. The spin polar angles right below saturation are
\begin{eqnarray}
& & \theta_i = \pi [ 1 + (-1)^i ] - (-1)^i \sin[\frac{\pi}{2N} (2i-1)] \theta_0
\end{eqnarray}
with $i=1,\dots,N$ and $\theta_0$ a very small parameter that goes to 0 at the ferromagnetic configuration (see App. \ref{appendix:saturationfield}). The sinusoidal dependence shows that the deviation from ferromagnetism is getting stronger going towards the center, with the influence of the field being the strongest close to the edges. For large $N$ the magnitude of the deviation is proportional to $\frac{\pi}{2N} (2i-1)$ for spins not very far from the edges, thus it increases linearly with the distance from the edges and differs proportionally to $\frac{\pi}{N}$ between successive spins when going towards the center.

\section{$J'=0$}
\label{sec:exchangeonly}

\begin{figure}
\includegraphics[width=3.5in,height=2.8in]{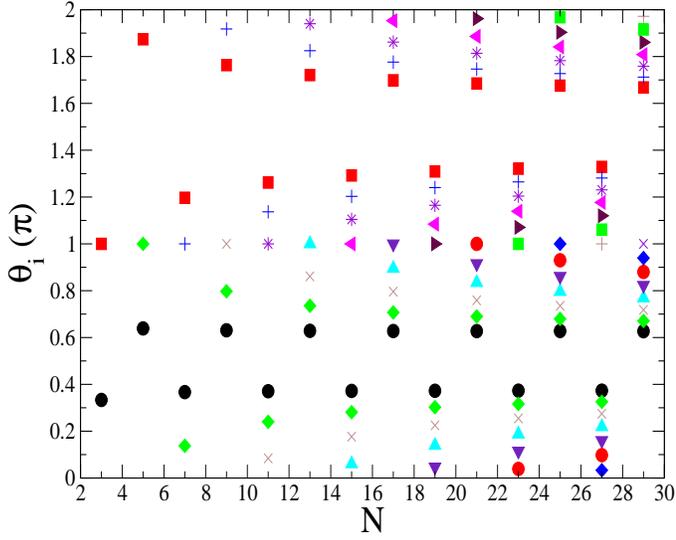}
\vspace{0pt}
\caption{Polar angles $\theta_i$ in units of $\pi$ for the $M=0$ lowest energy configuration as a function of the number of sites $N$ for $J'=0$. Different symbols denote different distance from the edge. Only spins from one side of the chain are shown as the symmetric ones have polar angles equal to $2\pi-\theta_i$. For $N=3$ the black circle corresponds to the edge spin and the red box to the central spin. For every subsequent odd $N$ a single angle is added, which always starts out at the center of the chain directed along $\pi$, has its own symbol, and its distance from the edge is $\frac{N-1}{2}$.
}
\label{fig:J=0S=0thetadistanceedge}
\end{figure}

For the smallest odd chain $N=3$ an analytic solution is possible by minimization of Hamiltonian (\ref{eq:Hchain}). According to Eqs (\ref{eq:thresholdfield}) and (\ref{eq:saturationfield}) it is $h_t=J$ and $h_s=3J$, and for this range of fields $M=\frac{h}{J} \geq M_{g,J' \leq \frac{J}{2}}=1$. The polar angles and ground state energy are ($\theta_3=\theta_1$ due to symmetry)
\begin{eqnarray}
\cos\theta_1 & = & \frac{M^2+3}{4M} \nonumber \\
\cos\theta_2 & = & \frac{M^2-3}{2M} \nonumber \\
E(M) & = & -\frac{J}{2}(5-M^2) .
\label{eq:N=3analytichigh}
\end{eqnarray}
When $M<M_{g,J' \leq \frac{J}{2}}$ analytic minimization of Hamiltonian (\ref{eq:Hchain2}) gives $M=\frac{J}{2G}$ with $G \geq \frac{J}{2}$, $\theta_3=2\pi-\theta_1$ and $\theta_2=\pi$, and
\begin{eqnarray}
\cos\theta_1 & = & \frac{M+1}{2} \nonumber \\
E(M) & = & -J(1+M) .
\end{eqnarray}
The dependence on $M$ is linear, in contrast to the quadratic dependence for $M \geq M_{g,J' \leq \frac{J}{2}}$ of Eqs (\ref{eq:N=3analytichigh}). It is ${\theta_1}(M=0)=\frac{\pi}{3}$ (Fig. \ref{fig:J=0S=0thetadistanceedge}), and the angle between the edge spins and the central spin is $\frac{2\pi}{3}$, thus the $M=0$ lowest energy configuration is the same with the ground state of the frustrated $N=3$ closed chain.

\begin{figure}
\includegraphics[width=3.5in,height=2.8in]{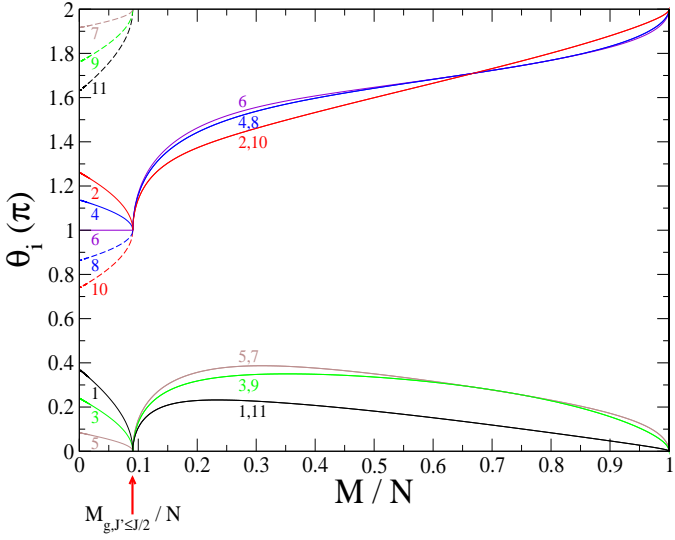}
\vspace{0pt}
\caption{Polar angles $\theta_i$ in units of $\pi$ for the lowest energy configuration as a function of the reduced total magnetization $M/N$ for $N=11$ and $J'=0$. The threshold field $h_t=0.285J$ and the saturation field $h_s=3.919J$ (Eqs (\ref{eq:thresholdfield}) and (\ref{eq:saturationfield})). The location of the residual magnetization $M_{g,J' \leq J/2}=1$ is shown with the (red) arrow. The indices (and colors) relate to the location of the spins with respect to the edge: $i=1, 11$: black, $i=2, 10$: red, $i=3, 9$: green, $i=4, 8$: blue, $i=5, 7$: brown, $i=6$: violet. For $M \geq M_{g,J' \leq J/2}$ the polar angles are symmetric with respect to the center of the chain. For $M < M_{g,J' \leq J/2}$ they add up to $2\pi$ for spins located symmetrically with respect to the center, and to show this the angles for the right half of the chain are indicated with long-dashed lines.
}
\label{fig:N=11J'=0polarangles}
\end{figure}

For $N>3$ the lowest energy configuration as a function of $M$ was calculated numerically. Results for $N=11$ are shown in Fig. \ref{fig:N=11J'=0polarangles}. For $M \geq M_{g,J' \leq \frac{J}{2}}=1$ the polar angles deviate from the AFM configuration to gain magnetic energy and their dependence on $M$ is not necessarily monotonic, similarly to the even $N$ case \cite{Rossler04}.
The odd chain behaves like a small magnet for small fields due to its residual magnetization, while the even chain has no residual magnetization. The odd spins practically point along an infinitesimal field and move away from it with increasing field the more the closer to the center they are, but eventually move back towards it as the magnetic energy becomes much stronger than the exchange energy for higher fields. The even spins are antiparallel to an infinitesimal field and vary monotonically with increasing field until they align themselves with it at saturation. Initially they deviate the more from the field direction the farther from the center they are. Similarly to even chains there exists a value of $M$ where all spins have the same deviation from the field direction except from the edge spins, coined knot point in Ref. \cite{Rossler04}, where the relative deviation from the $x$ axis changes order for the even spins, but not for the odd spins. For even chains the relative deviation changes order at the knot point for both even and odd spins. Fig. \ref{fig:N=11configurations}(b) shows the spin configuration for the $M=3$ ground state, and Fig. \ref{fig:reducedmagnetization} the reduced magnetization $\frac{M}{N}$ as a function of the external magnetic field $h$ over its saturation value $h_{sat}$ (Eq. (\ref{eq:saturationfield})) for $J'=0$. For $h$ less than the threshold field $h_t$ of Eq. (\ref{eq:thresholdfield}) the magnetization is equal to the zero field residual magnetization $M_{g,J' \leq \frac{J}{2}}=1$.

When $M<M_{g,J' \leq \frac{J}{2}}$ the central spin is fixed along the $\pi$ direction (Fig. \ref{fig:N=11J'=0polarangles}). Spins that come in pairs sharing the polar angles for $M \geq M_{g,J' \leq \frac{J}{2}}$ now split up symmetrically with respect to the $x$ axis, while the deviation from the $x$ axis of the spins within the two groups which have values less and greater than $\pi$ reverses its order in comparison to the $M \geq M_{g,J' \leq \frac{J}{2}}$ case before the knot point. Thus spins closer to the center deviate the least from their directions in the AFM configuration. Fig. \ref{fig:N=11configurations}(a) shows the spins in the $M=0$ lowest energy configuration.

The polar angles for the lowest energy configuration with $M=0$ are plotted in Fig. \ref{fig:J=0S=0thetadistanceedge} for varying $N$. The central spin lies along the $\pi$ direction. The angles tend to limiting values as functions of $N$, under the constraint that for two adjacent sizes the values of the polar angle of a spin at a fixed distance from the edge tend to become supplementary modulo $2\pi$ as $N \to \infty$. Convergence of the angles with $N$ is faster going towards the edges. Spins tend to be antiparallel with increasing $N$ the closer to the edge they are, when counting of pairs starts from the second and third spin from the edge.

\section{$J' \neq 0$}
\label{sec:exchangewithbiquadratic}

When $N=3$ according to Eqs (\ref{eq:thresholdfield}) and (\ref{eq:saturationfield}) it is $h_t=J-2J'$ and $h_s=3(J+2J')$.
An analytic solution can now be found for $M_{g,J' > \frac{J}{2}}^{min}=|1-\frac{J}{J'}|$.
For $M<M_{g,J' \leq \frac{J}{2}}$ the lowest energy configuration of Hamiltonian (\ref{eq:Hchain2}) has $M=\frac{J-J'}{2G+J'}$ with $G \geq \frac{J}{2}-J'$, and like the $J'=0$ case $\theta_3=2\pi-\theta_1$ and $\theta_2=\pi$, with $\cos\theta_1=\frac{M+1}{2}$. The energy is $E(M)=-(J-\frac{1+M}{2}J')(1+M)$.

\begin{figure}
\includegraphics[width=3.5in,height=2.8in]{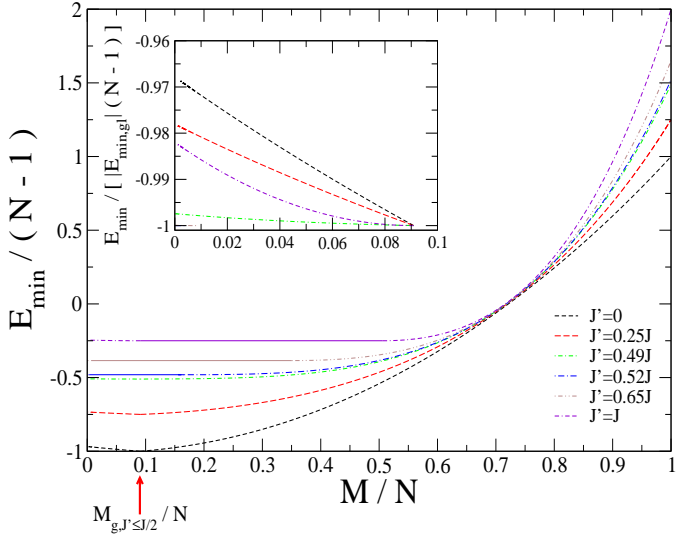}
\vspace{0pt}
\caption{Lowest energy per bond as a function of the reduced magnetization $M/N$ for $N=11$ and different $\frac{J'}{J}$. For $\frac{J'}{J} \leq \frac{1}{2}$ it is $M_{g,J' \leq J/2}=1$ (shown with the (red) arrow) for the global ground state, while for $\frac{J'}{J} > \frac{1}{2}$ the global ground state corresponds to magnetization values ranging from $M_{g,J'>J/2}^{min}$ to $M_{g,J'>J/2}^{max}$ (Fig. \ref{fig:redmagngrstateminmax}), highlighted with solid lines. The inset focuses on magnetization values less than the minimum of the global ground state, with the energies divided by the absolute value of the global energy minimum. For $\frac{J'}{J}=0.52$ and 0.65 the magnetization range is much smaller and the corresponding reduced energies are very close to -1.
}
\label{fig:lowestenergieswithS}
\end{figure}

For $N>3$ numerical results are presented again for $N=11$. Fig. \ref{fig:lowestenergieswithS} shows the lowest energy per bond as a function of $M$ for different $\frac{J'}{J}$. The energy decreases in magnitude with increasing $\frac{J'}{J}$ due to the stronger competition between the biquadratic and the bilinear exchange. For $J' > \frac{J}{2}$ the lowest energy corresponds to the ground state manifold that has $M$ between $M_{g,J' > \frac{J}{2}}^{min}$ and $M_{g,J' > \frac{J}{2}}^{max}$ (Sec. \ref{subsec:minimummaximummagnetization}). For $\frac{M}{N} \sim 0.75$ all the energy curves are close to zero, the reason being that nearest-neighbor spins are very close to normal.

\begin{figure}
\includegraphics[width=3.5in,height=2.8in]{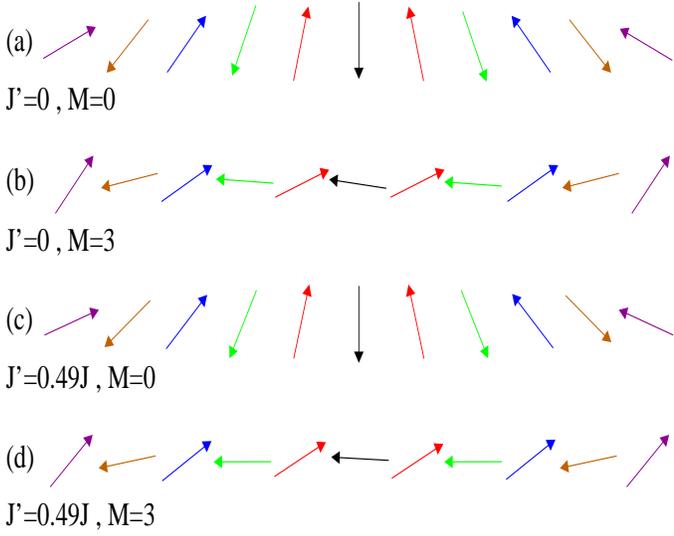}
\vspace{0pt}
\caption{Lowest energy configurations for $N=11$, $J'=0$, and total magnetization (a) $M=0$, and (b) $M=3$. The corresponding configurations for $J'=0.49J$ are shown in (c) and (d). Spins located symmetrically with respect to the center are shown with the same color, with the color coding following Figs \ref{fig:N=11J'=0polarangles} and \ref{fig:N=11J'=0.49polarangles}. For $M=3$ the polar angles are symmetric with respect to the center of the chain. For $M=0$ they add up to $2\pi$ for spins located symmetrically with respect to the center. $M$ is opposite in direction to the central spin of (a) and (c).
}
\label{fig:N=11configurations}
\end{figure}

\begin{figure}
\includegraphics[width=3.5in,height=2.8in]{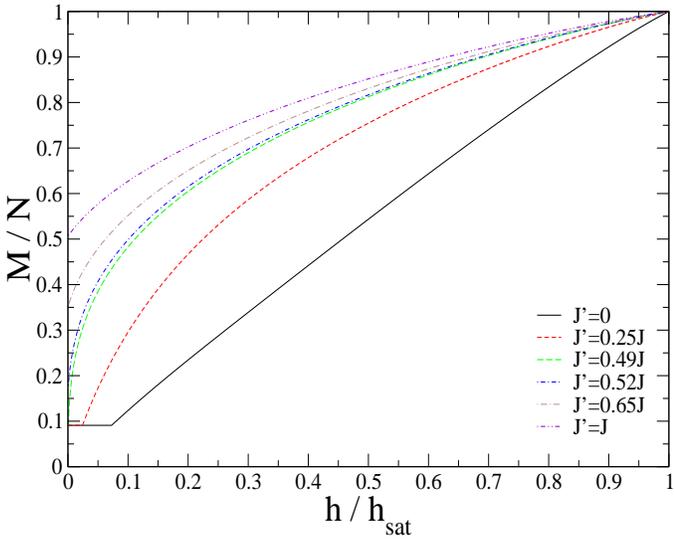}
\vspace{0pt}
\caption{Reduced magnetization $\frac{M}{N}$ as a function of the magnetic field $h$ over its saturation value $h_{sat}$ (Eq. (\ref{eq:saturationfield})) for $N=11$ and different $\frac{J'}{J}$. For $\frac{J'}{J} \leq \frac{1}{2}$ the threshold field given by Eq. (\ref{eq:thresholdfield}) is required to destroy the AFM configuration, while for $\frac{J'}{J} > \frac{1}{2}$ an infinitesimal field picks the global ground state with maximum residual magnetization given by Eq. (\ref{eq:resspinmax}).
}
\label{fig:reducedmagnetization}
\end{figure}

Fig. \ref{fig:N=11J'=0.49polarangles} shows the polar angles in the lowest energy configuration for $N=11$ as a function of $\frac{M}{N}$ for $\frac{J'}{J}=0.49$. In comparison with $J'=0$ (Fig. \ref{fig:N=11J'=0polarangles}) nearest-neighbor polar angles do not differ much for $M \geq M_{g,J' \leq \frac{J}{2}}$ even though now the biquadratic is strongly competing with the bilinear exchange interaction, bearing in mind the differences in the threshold and saturation field (Sec. \ref{subsec:thresholdfield} and \ref{subsec:saturationmagneticfield}). This can also be seen by comparing the $M=3$ ground state configuration with the corresponding one for $J'=0$ (Fig. \ref{fig:N=11configurations}(d) and (b) respectively). Still it takes a larger $\frac{M}{N}$ to reach the knot point, where all spin deviations from the field direction but the ones at the edges are equal. One can think of $M$ as generated by the external magnetic field that now has to compensate also for the biquadratic energy to drive the system towards saturation (refer to Eq. (\ref{eq:saturationfield}) for the saturation field where $J$ and $J'$ add up). Fig. \ref{fig:reducedmagnetization} shows the reduced magnetization in a field for $\frac{J'}{J}=0.49$. Due to the non-zero biquadratic interaction the threshold field $h_t$ given by Eq. (\ref{eq:thresholdfield}) is much smaller in comparison with the $J'=0$ case, while the saturation field given by Eq. (\ref{eq:saturationfield}) increases. The susceptibility is less uniform in comparison with the $J'=0$ case. For $M < M_{g,J' \leq \frac{J}{2}}$ the angles in Fig. \ref{fig:N=11J'=0.49polarangles} are more spread out within the three different groups in comparison with $J'=0$. Still the polar angles are very similar, as can also be seen by comparing the $M=0$ lowest energy configurations (Fig. \ref{fig:N=11configurations}(c) and (a)).

\begin{figure}
\includegraphics[width=3.5in,height=2.8in]{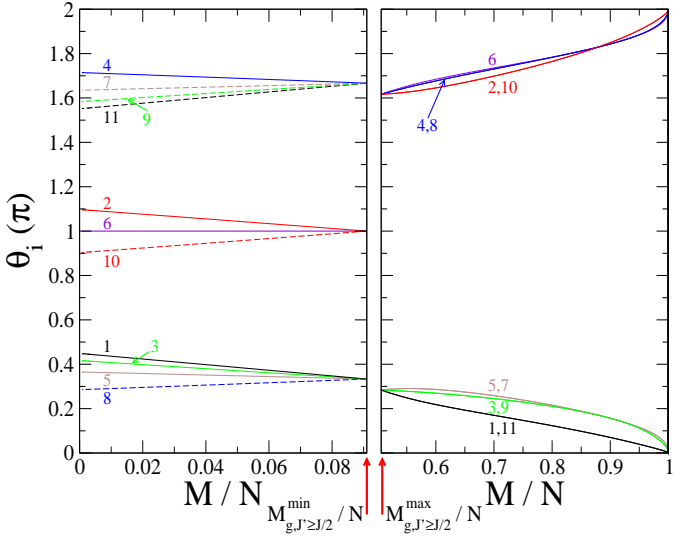}
\vspace{0pt}
\caption{Polar angles $\theta_i$ in units of $\pi$ for the lowest energy configuration as a function of the reduced total magnetization $M/N$ for $N=11$ and $J'/J=1$. The saturation field $h_s=11.757 J$ (Eq. (\ref{eq:saturationfield})). The minimum and maximum values of the residual magnetization are $M_{g,J' > J/2}^{min}=1$ and $M_{g,J' > J/2}^{max}=\sqrt{31}$ (Eq. (\ref{eq:resspinmax})), and are shown with red arrows. The plot is for total magnetizations outside this range. The indices (and colors) relate to the location of the spins with respect to the edge: $i=1, 11$: black, $i=2, 10$: red, $i=3, 9$: green, $i=4, 8$: blue, $i=5, 7$: brown, $i=6$: violet. For $M \geq M_{g,J' > J/2}^{max}$ the polar angles are symmetric with respect to the center of the chain. For $M \leq M_{g,J' > J/2}^{min}$ they add up to $2\pi$ for spins located symmetrically with respect to the center, and to show this the angles for the right half of the chain are indicated with long-dashed lines.
}
\label{fig:N=11J'=1polarangles}
\end{figure}

\begin{figure}
\includegraphics[width=3.5in,height=2.8in]{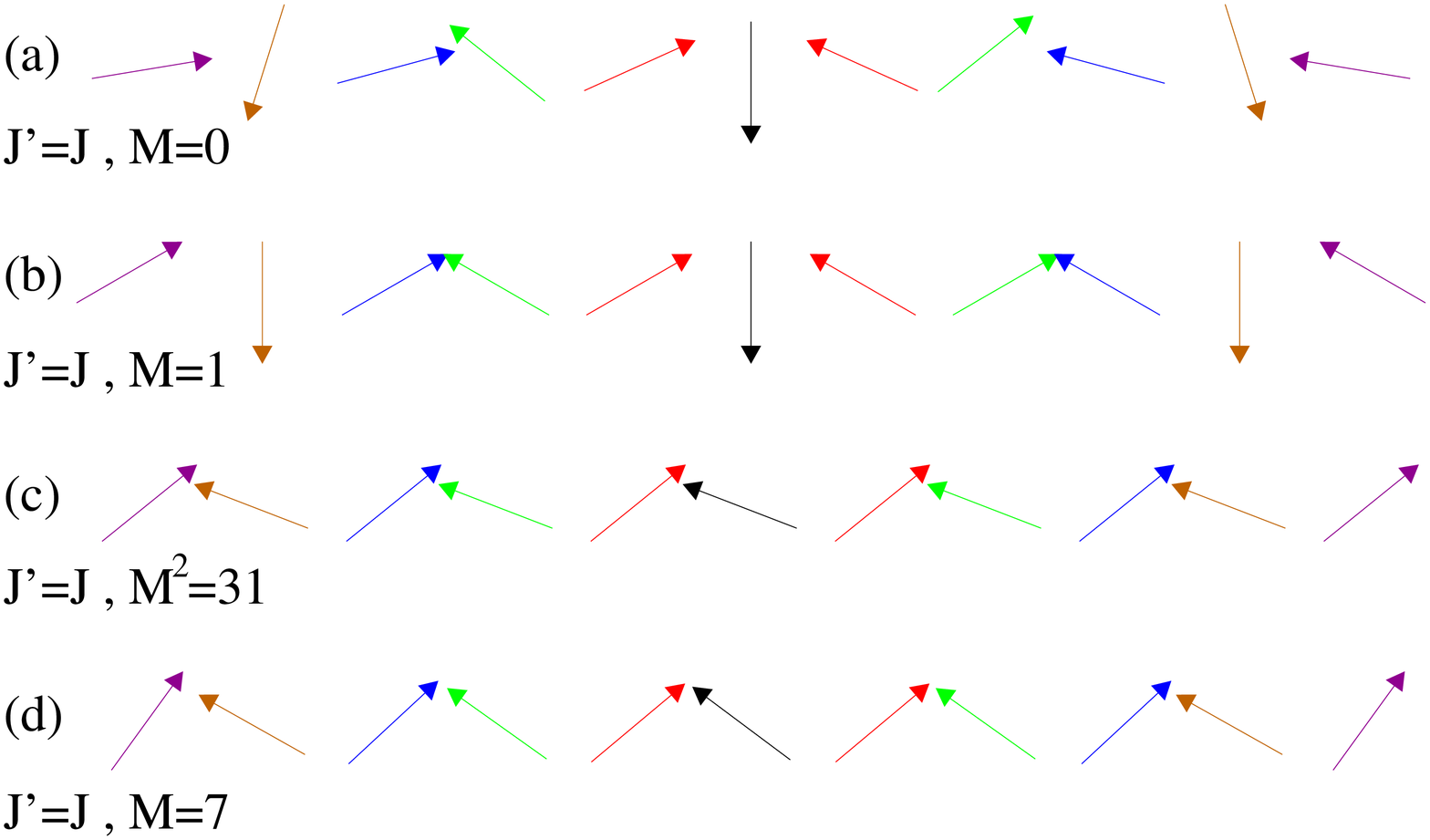}
\vspace{0pt}
\caption{Lowest energy configurations for $N=11$, $J'=J$, and total magnetization (a) $M=0$, (b) $M=1$, (c) $M=\sqrt{31}$, and (d) $M=7$. Spins located symmetrically with respect to the center are shown with the same color, with the color coding following Fig. \ref{fig:N=11J'=1polarangles}. For $M=\sqrt{31}$ and 7 the polar angles are symmetric with respect to the center of the chain. For $M=0$ and 1 they add up to $2\pi$ for spins located symmetrically with respect to the center. $M$ is opposite in direction to the central spin of (a) and (b). For M=1 other degenerate configurations are also possible.
}
\label{fig:N=11configurationsJprime=1}
\end{figure}

In Fig. \ref{fig:N=11J'=1polarangles} the polar angles in the lowest energy configuration for $N=11$ are plotted as a function of $\frac{M}{N}$ for $\frac{J'}{J}=1$. The ground state is continuously degenerate and corresponds to magnetization values ranging from $M_{g,J' > \frac{J}{2}}^{min}=1$ to $M_{g,J' > \frac{J}{2}}^{max}=\sqrt{31}$ (Eq. (\ref{eq:resspinmax})), and the polar angles are plotted for magnetizations outside this range. In comparison with Figs \ref{fig:N=11J'=0polarangles} and \ref{fig:N=11J'=0.49polarangles} and due to the reduced magnetization range the polar angles are monotonic for $M \geq M_{g,J' > \frac{J}{2}}^{max}$. For $M \leq M_{g,J' > \frac{J}{2}}^{min}$
spins 4 and 8 have left the middle group where they belonged to for $J' \leq \frac{J}{2}$ and are now the spins closest to the $x$ axis among all spins. Fig. \ref{fig:N=11configurationsJprime=1} depicts the lowest energy configurations for different values of $M$. The configuration does not change very much between $M=\sqrt{31}$ and 7. For $M=1$ the angle between nearest-neighbors is constant and equal to $\frac{2\pi}{3}$, while the $M=0$ configuration is considerably different from the ones for $\frac{J'}{J} \leq \frac{1}{2}$ shown in Fig. \ref{fig:N=11configurations}(a) and (c). Fig. \ref{fig:reducedmagnetization} shows the magnetization curve in a field also for $\frac{J'}{J} > \frac{1}{2}$. An infinitesimal field picks the global ground state with maximum residual magnetization given by Eq. (\ref{eq:resspinmax}).

\section{Conclusions}
\label{sec:conclusions}
The lowest energy configuration of the classical spin model with AFM bilinear and biquadratic exchange interactions has been calculated for small odd chains. The odd chains differ from their even counterparts in that they have a residual magnetization for zero field, therefore they act like small magnets for small fields. This results in a threshold magnetic field that needs to be applied to divert the spins from the AFM configuration, making the latter robust to weak fields in contrast to even chains. Furthermore, the lowest energy configuration for $M$ less than the (minimum) value of the residual magnetization in zero field is not accessible with a magnetic field. The lowest energy and the corresponding spin configuration as functions of $M$ depend on the relative ratio of the biquadratic to the bilinear exchange interaction. They have been calculated for the whole range of $M$, even when $M$ is smaller than the (minimum) value of the global energy minimum. Analytic expressions were derived for the threshold and saturation field and the polar angles in their vicinity, as well as for the maximum residual magnetization in the absence of a field for relatively stronger biquadratic interaction.

Since an odd open chain has an absolute ground state with finite spin, it has potential for magnetic storage. Logic gates based on magnetic nanochains fabricated on surfaces with STM have already been built, and the parity of the nanochain determines the logical operation of the device \cite{Khajetoorians11}. What is important for these logic gates is that the individual magnetic atoms function as two state (up and down magnetization) bits, something effected by a strong magnetic anisotropy. In the present paper it has been shown that the absolute ground state of the classical AFM Heisenberg model acquires a richer structure with the inclusion of biquadratic exchange interactions. While the absolute ground state is AFM for smaller biquadratic exchange, for stronger it corresponds to a range of magnetization values where the configuration with maximum magnetization can be selected with an infinitesimal magnetic field. This also affects the accesible magnetization values with a finite external field. The biquadratic exchange in odd chains could thus provide a means to control the magnitude of the total magnetization, as well as its distribution among the spins. Since classical spins provide a very good approximation of the behavior of the chain down to relatively small spin magnitudes $s$ \cite{Machens13,NPK13}, the findings in this paper show that apart from extra spins attached to the chain its magnetic properties can be manipulated by additional interaction terms allowed from the theoretical point of view.

\begin{appendix}

\section{Maximum Residual Magnetization of the Ground State Manifold for $J' > \frac{J}{2}$}
\label{appendix:limitingmagnetizationfromabove}
An infinitesimally small magnetic field picks out from the $J' > \frac{J}{2}$ degenerate zero field configurations (Sec. \ref{subsec:minimummaximummagnetization}) the one with maximum residual magnetization $M_{g,J' > \frac{J}{2}}^{max}$. Nearest-neighbor spins are at a relative angle $\theta_g=\arccos(-\frac{J}{2J'})$, and to maximize $M_{g,J' > \frac{J}{2}}$ (and the magnetic energy) all spins must lie in a plane that includes the infinitesimal magnetic field. If the first spin is at an angle $\theta_0 \leq \frac{\pi}{2}$ with respect to the field direction, the second will be at an angle $\theta_g-\theta_0$. It is $\frac{\pi}{2} \leq \theta_g < \pi$, therefore to maximize $M_{g,J' > \frac{J}{2}}$ and avoid any spins with positive magnetic energy all subsequent spins will successively be directed at these two angles. Thus there are $\frac{N+1}{2}$ spins at an angle $\theta_0$ with the field and  $\frac{N-1}{2}$ spins at an angle $\theta_g-\theta_0$. The maximum residual magnetization is
\begin{eqnarray}
M_{g,J' > \frac{J}{2}}^{max} = \frac{N+1}{2} \cos\theta_0 + \frac{N-1}{2} \cos(\theta_g-\theta_0) .
\label{eq:hamiltmaxresmagn}
\end{eqnarray}
Taking the derivative with respect to $\theta_0$ and setting it to zero (which also guarantees that the magnetization perpendicular to the field is zero), it is after some algebra
\begin{eqnarray}
\theta_0 = \arctan\frac{\sin\theta_g}{\frac{N+1}{N-1} + \cos\theta_g} .
\end{eqnarray}
Since $\frac{\pi}{2} \leq \theta_g < \pi$ it is $\sin\theta_g>0$, thus $\sin\theta_g=\sqrt{1-\cos^2\theta_g}\Rightarrow \sin\theta_g=\sqrt{1-\frac{J^2}{4J'^2}}$. Then
\begin{eqnarray}
\theta_0 = \arctan\frac{\sqrt{1-\frac{J^2}{4J'^2}}}{\frac{N+1}{N-1} - \frac{J}{2J'}} .
\label{eq:appb1}
\end{eqnarray}
It is $\theta_0(J' \gg J)=\arctan\frac{N-1}{N+1}$. For an infinite chain $\theta_0(J' \gg J,N \to \infty) \to \frac{\pi}{4}$.

The numerator and denominator in Eq. (\ref{eq:appb1}) are both positive, therefore $\theta_0 \leq \frac{\pi}{2}$ as it should be to maximize the magnetic energy. Then it is
\begin{eqnarray}
\cos\theta_0 = \frac{1}{\sqrt{1+\tan^2\theta_0}} \Rightarrow \cos\theta_0 = \frac{1}{\sqrt{1+\frac{1-\frac{J^2}{4J'^2}}{(\frac{N+1}{N-1} - \frac{J}{2J'})^2}}} .
\end{eqnarray}
Similarly
\begin{eqnarray}
\sin\theta_0 = \frac{\tan\theta_0}{\sqrt{1+\tan^2\theta_0}} \Rightarrow \sin\theta_0 = \frac{\frac{\sqrt{1-\frac{J^2}{4J'^2}}}{\frac{N+1}{N-1} - \frac{J}{2J'}}}{\sqrt{1+\frac{1-\frac{J^2}{4J'^2}}{(\frac{N+1}{N-1} - \frac{J}{2J'})^2}}} .
\end{eqnarray}
After some algebra (\ref{eq:hamiltmaxresmagn}) gives
\begin{eqnarray}
M_{g,J'>\frac{J}{2}}^{max} & = & \frac{N+1}{2} \cos\theta_0 + \nonumber \\ & & \frac{N-1}{2} ( \cos\theta_g \cos\theta_0 + \sin\theta_g \sin\theta_0 ) \Rightarrow \nonumber \\ M_{g,J'>\frac{J}{2}}^{max} & = & \frac{1}{2\sqrt{1+\frac{1-\frac{J^2}{4J'^2}}{(\frac{N+1}{N-1}-\frac{J}{2J'})^2}}} [ N+1+(N-1) \nonumber \\ & & (-\frac{J}{2J'} + \frac{1-\frac{J^2}{4J'^2}}{\frac{N+1}{N-1} - \frac{J}{2J'}}) ] .
\end{eqnarray}
For an infinite chain $\frac{M_{g,J' > \frac{J}{2},N \to \infty}^{max}}{N} = \sqrt{\frac{1-\frac{J}{2J'}}{2}}$. Then $\frac{M_{g,N \to \infty,J' \to \infty}^{max}}{N} = \frac{1}{\sqrt{2}}$. When $J' \to \infty$ it is $M_{g,J' \to \infty}^{max} = \sqrt{\frac{N^2+1}{2}}$.

\section{Threshold Magnetic Field for $J' \leq \frac{J}{2}$}
\label{appendix:thresholdfield}

An odd chain with length $N$ in a magnetic field maintains its AFM configuration up to a threshold field $h_t$. For $h \leq h_t$ odd spins have $\theta_i=0$, $i=1,3,\dots,N$, while even spins $\theta_i=\pi$, $i=2,4,\dots,N-1$.
Hamiltonian (\ref{eq:Hchain}) of the main paper can be rewritten in polar coordinates for finite $h$, and then expressed with appropriate indices with respect to the chain's even and odd sites. If for every even spin the transformation $\theta_{2i} \to \theta_{2i} - \pi$ is performed due to the low field AFM configuration, then if the spins start to slightly tilt away from it a small angle expansion of the Hamiltonian gives

\begin{eqnarray}
H & = & -(N-1)(J-J') - h + \frac{1}{2} (J-2J') \nonumber \\ & & \sum_{i=1}^{\frac{N-1}{2}} [ (\theta_{2i-1}-\theta_{2i})^2 + ( \theta_{2i} - \theta_{2i+1} )^2 ] + \nonumber \\ & & \frac{h}{2} \sum_{i=1}^{\frac{N-1}{2}} ( \theta_{2i-1}^2 - \theta_{2i}^2 ) + \frac{h}{2} \theta_N^2 .
\label{eq:hamiltthresholdsmallangle}
\end{eqnarray}
The derivatives of Hamiltonian (\ref{eq:hamiltthresholdsmallangle}) with respect to the $\theta_i$ are:
\begin{eqnarray}
\frac{\partial H}{\partial \theta_1} & = & (J-2J') (\theta_1 - \theta_2) + h \theta_1 \nonumber \\
\frac{\partial H}{\partial \theta_{2i}} & = & (J-2J') (2 \theta_{2i} - \theta_{2i+1} - \theta_{2i-1}) - h \theta_{2i} \textrm{ , } \nonumber \\ & & i=1,\dots,\frac{N-1}{2} \nonumber \\
\frac{\partial H}{\partial \theta_{2i-1}} & = & (J-2J') (2 \theta_{2i-1} - \theta_{2i} - \theta_{2i-2}) + h \theta_{2i-1} \textrm{ , } \nonumber \\ & & i=2,\dots,\frac{N-1}{2} \nonumber \\
\frac{\partial H}{\partial \theta_N} & = & (J-2J') (\theta_N - \theta_{N-1}) + h \theta_N .
\label{eq:hamiltthresholdderivatives}
\end{eqnarray}
To get the minima the derivatives (\ref{eq:hamiltthresholdderivatives}) are set equal to 0. To simplify the expressions $\alpha \equiv J-2J'$. Then Eqns (\ref{eq:hamiltthresholdderivatives}) define the following system:

\begin{widetext}
\begin{displaymath}
\left(
\begin{array}{ccccccccccc}
\alpha+h & -\alpha & 0 & 0 & 0 & \ldots & 0 & 0 & 0 & 0 & 0 \\
-\alpha & 2\alpha-h & -\alpha & 0 & 0 & \ldots & 0 & 0 & 0 & 0 & 0 \\
0 & -\alpha & 2\alpha+h & -\alpha & 0 & \ldots & 0 & 0 & 0 & 0 & 0 \\
\vdots & \vdots & \vdots & \vdots & \vdots & \ddots & \vdots & \vdots & \vdots & \vdots & \vdots \\
0 & 0 & 0 & 0 & 0 & \ldots & 0 & -\alpha & 2\alpha+h & -\alpha & 0 \\
0 & 0 & 0 & 0 & 0 & \ldots & 0 & 0 & -\alpha & 2\alpha-h & -\alpha \\
0 & 0 & 0 & 0 & 0 & \ldots & 0 & 0 & 0 & -\alpha & \alpha+h
\end{array} \right)
\left(
\begin{array}{c}
\theta_1 \\
\theta_2 \\
\theta_3 \\
\vdots \\
\theta_{N-2} \\
\theta_{N-1}\\
\theta_{N} \\
\end{array} \right) =
\left(
\begin{array}{c}
0 \\
0 \\
0 \\
\vdots \\
0 \\
0 \\
0 \\
\end{array} \right) .
\end{displaymath}
\end{widetext}

The system of equations is homogeneous and has a solution if the determinant of the matrix is zero. The characteristic polynomial of the matrix is \cite{Kouachi08}
\begin{eqnarray}
\Delta_n & = & \frac{\alpha^{N-1}}{\sin(2\theta)} \{(h-\lambda)\sin[(N+1)\theta]- \nonumber \\ & & (h+\lambda)\sin[(N-1)\theta]\}
\label{eq:hamiltthresholddeterminant}
\end{eqnarray}
where $\lambda$ is an eigenvalue of the matrix and
\begin{eqnarray}
(2\alpha+h-\lambda)(2\alpha-h-\lambda)=4\alpha^2\cos^2\theta .
\label{eq:hamiltthresholddeterminant1}
\end{eqnarray}
To calculate the determinant $\lambda=0$ must be taken, and (\ref{eq:hamiltthresholddeterminant}) gives
\begin{eqnarray}
\Delta_n (\lambda=0) & = & \frac{\alpha^{N-1}h}{\sin(2\theta)} \{\sin[(N+1)\theta] - \nonumber \\ & & \sin[(N-1)\theta]\}
\end{eqnarray}
and from (\ref{eq:hamiltthresholddeterminant1}) it is $h_t=2 \alpha \sin\theta$.
Taking into account that $\sin[(N+1) \theta]=2\cos\theta \sin(N\theta)-\sin[(N-1)\theta]$ it is eventually

\begin{eqnarray}
\Delta_n(\lambda=0) & = & \frac{\alpha^{N-1}h\cos[(N-1)\theta]}{\sin(2\theta)} \{ \sin(2\theta) - [1-\cos(2\theta)] \nonumber \\ & & \tan[(N-1)\theta] \} .
\end{eqnarray}

After some algebra $\Delta_n(\lambda=0)=0$ implies that $\tan[(N-1)\theta] = \cot\theta$
or $\theta = \frac{\pi}{2N}$. Then it is $h_t = 2\alpha \sin\frac{\pi}{2N}$ or (given in \cite{Politi09} for $J'=0$)
\begin{eqnarray}
h_t = 2(J-2J') \sin\frac{\pi}{2N} .
\end{eqnarray}
$h_t$ is a monotonic function of $N$ with $\lim_{N \to \infty} h_t=0$. The eigenvector that corresponds to $\lambda=0$ and $\theta=\frac{\pi}{2N}$ is given by \cite{Kouachi08}
\begin{eqnarray}
\theta_i & = & (-\alpha)^N \{ \alpha \sin[\frac{\pi}{2N} (N-i+2)] - \nonumber \\ & & (\alpha-h_t) \sin[\frac{\pi}{2N} (N-i)] \}\textrm {, i odd} \nonumber \\
\theta_i & = & (-\alpha)^N \{ (\alpha+h_t) \sin[\frac{\pi}{2N} (N-i+1)] - \nonumber \\ & & \alpha \sin[\frac{\pi}{2N} (N-i-1)] \}\textrm {, i even}
\end{eqnarray}
with $i=1,\dots,N$. After some algebra, replacing $h_t$,
defining the constant $2 (-\alpha)^{N+1} \sin\frac{\pi}{2N} \equiv C_N$
and taking into account that the even spins have undergone the transformation $\theta_{2i} \to \theta_{2i} - \pi$, it finally is
\begin{eqnarray}
\theta_i & = & C_N \{ \sin[\frac{\pi}{2N}(i-1)] + \cos(\frac{\pi}{2N}i) \}\textrm {, i odd} \nonumber \\
\theta_i & = & \pi + C_N \{ \sin(\frac{\pi}{2N}i) + \cos[\frac{\pi}{2N}(i-1)] \} \textrm {,} \nonumber \\ & & \textrm {i even} .
\end{eqnarray}
By substituting $N-i+1$ for $i$ it is straightforward to show that these expressions are symmetric with respect to the center of the chain.

\section{Saturation Magnetic Field}
\label{appendix:saturationfield}

The following derivation does not depend on the parity of $N$ and is thus also valid for even chains.
Hamiltonian (\ref{eq:Hchain}) of the main paper can be expressed in polar coordinates for finite $h$. Very close to saturation the odd polar angles $\theta_i$, $i=1,3,\dots,N$ are very small. The even angles $\theta_i$, $i=2,4,\dots,N-1$ are very close to $2\pi$, therefore the transformation $\theta_{2i} \to 2\pi - \theta_{2i}$ can be performed. If the spins start to slightly tilt away from their ferromagnetic configuration a small angle expansion of the Hamiltonian leads to, since every bond has one even spin

\begin{eqnarray}
H & = & (N-1)(J+J') -N h - \frac{1}{2} (J+2J') \nonumber \\ & & \sum_{i=1}^{N-1} (\theta_i + \theta_{i+1})^2 + \frac{h}{2} \sum_{i=1}^N \theta_i^2 .
\label{eq:appd3}
\end{eqnarray}
The derivatives of Hamiltonian (\ref{eq:appd3}) with respect to the $\theta_i$ are
\begin{eqnarray}
\partial H / \partial \theta_1 & = & -(J+2J') ( \theta_1 + \theta_2 ) + h \theta_1 \nonumber \\
\partial H / \partial \theta_i & = & -(J+2J') ( 2 \theta_i + \theta_{i+1} + \theta_{i-1} ) + h \theta_i \nonumber \\ & & i=2,\dots,N-1 \nonumber \\
\partial H / \partial \theta_N & = & -(J+2J') ( \theta_N + \theta_{N-1} ) + h \theta_N .
\label{eq:appd4}
\end{eqnarray}

To get the minima the derivatives (\ref{eq:appd4}) are set equal to 0. To simplify the expressions $\alpha \equiv J+2J'$. Then Eqns (\ref{eq:appd4}) define the following system:
\begin{widetext}
\begin{displaymath}
\left(
\begin{array}{ccccccccccc}
\alpha-h & \alpha & 0 & 0 & 0 & \ldots & 0 & 0 & 0 & 0 & 0 \\
\alpha & 2\alpha-h & \alpha & 0 & 0 & \ldots & 0 & 0 & 0 & 0 & 0 \\
0 & \alpha & 2\alpha-h & \alpha & 0 & \ldots & 0 & 0 & 0 & 0 & 0 \\
\vdots & \vdots & \vdots & \vdots & \vdots & \ddots & \vdots & \vdots & \vdots & \vdots & \vdots \\
0 & 0 & 0 & 0 & 0 & \ldots & 0 & \alpha & 2\alpha-h & \alpha & 0 \\
0 & 0 & 0 & 0 & 0 & \ldots & 0 & 0 & \alpha & 2\alpha-h & \alpha \\
0 & 0 & 0 & 0 & 0 & \ldots & 0 & 0 & 0 & \alpha & \alpha-h
\end{array} \right)
\left(
\begin{array}{c}
\theta_1 \\
\theta_2 \\
\theta_3 \\
\vdots \\
\theta_{N-2} \\
\theta_{N-1}\\
\theta_{N} \\
\end{array} \right) =
\left(
\begin{array}{c}
0 \\
0 \\
0 \\
\vdots \\
0 \\
0 \\
0 \\
\end{array} \right) .
\end{displaymath}
\end{widetext}
The system of equations is homogeneous and has a solution if the determinant of the matrix is zero. One has to calculate the eigenvalues of the matrix and pick out the zero eigenvalue. The eigenvalues are
after substituting $\alpha$ \cite{Lombardi88}
\begin{eqnarray}
\lambda_s = 2(J+2J')-h+2(J+2J')\cos\frac{k \pi}{N}, \textrm { }k=1,\dots,N . \nonumber
\end{eqnarray}

Setting $\lambda_s=0$ it is
\begin{eqnarray}
h' = 2(J+2J')(1+\cos\frac{\pi k}{N}), \textrm { }k=1,\dots,N . \nonumber
\end{eqnarray}
The maximum value for $h'$ is generated when the argument of the cosine is minimum, and this corresponds to $k=1$, therefore
\begin{eqnarray}
& & h_s = 2(J+2J') (1+\cos\frac{\pi}{N}) .
\end{eqnarray}
$h_s$ is a monotonic function of $N$ with $\lim_{N \to \infty} h_s=4(J+2J')$. The corresponding normalized eigenvector for $k=1$ is \cite{Lombardi88}
\begin{eqnarray}
\theta_i = \sqrt{\frac{2}{N}} \sin[\frac{\pi}{2N} (2i-1)], \textrm { }i=1,\dots,N . \nonumber
\end{eqnarray}
Defining the constant $\sqrt{\frac{2}{N}} \equiv C_N$
and taking into account that the even spins have undergone the transformation $\theta_{2i} \to 2\pi - \theta_{2i}$, it finally is
\begin{eqnarray}
\theta_i = \pi[1+(-1)^i]-(-1)^i C_N \sin[\frac{\pi}{2N} (2i-1)], \textrm { }i=1,\dots,N . \nonumber
\end{eqnarray}

By substituting $N-i+1$ for $i$ it is straightforward to show that the expression is symmetric with respect to the center of the chain.

\end{appendix}

\bibliography{paperoddchains}

\newpage
\newpage

\vspace{20pt}

\begin{figure}
\includegraphics[width=3.5in,height=2.8in]{paperoddchainsfig4.eps}
\vspace{0pt}
\caption{Polar angles $\theta_i$ in units of $\pi$ for the lowest energy configuration as a function of the reduced total magnetization $M/N$ for $N=11$ and $J'/J=0.49$. The threshold field $h_t=5.693 \times 10^{-3} J$ and the saturation field $h_s=7.760 J$ (Eqs (\ref{eq:thresholdfield}) and (\ref{eq:saturationfield})). The location of the residual magnetization $M_{g,J' \leq J/2}=1$ is shown with the (red) arrow. The indices (and colors) relate to the location of the spins with respect to the edge: $i=1, 11$: black, $i=2, 10$: red, $i=3, 9$: green, $i=4, 8$: blue, $i=5, 7$: brown, $i=6$: violet. For $M \geq M_{g,J' \leq J/2}$ the polar angles are symmetric with respect to the center of the chain. For $M < M_{g,J' \leq J/2}$ they add up to $2\pi$ for spins located symmetrically with respect to the center, and to show this the angles for the right half of the chain are indicated with long-dashed lines.
}
\label{fig:N=11J'=0.49polarangles}
\end{figure}

\end{document}